\documentclass[11pt]{article}
\usepackage[a4paper,margin=2.5cm]{geometry}
\usepackage{amsmath,amssymb}
\usepackage{graphicx}
\usepackage{hyperref}
\usepackage{authblk}
\usepackage{setspace}
\usepackage{microtype}

\title{\textbf{Mandelbrot, Financial Markets and \\ the Origins of ``Econophysics''}}
\author{JP Bouchaud}
\affil{CFM \& Acad\'emie des Sciences}
\date{}

\begin{document}
\maketitle
\begin{abstract}
 This text\footnote{written for the proceedings of the conference 'Mandelbrot’s propositions in economics and finance, and their legacy', 10-11 December 2024, Maison des Sciences Économiques, Paris. I thank Christian Walter for inviting me and insisting I should write this piece.} revisits the origins of econophysics through the figure of Benoît Mandelbrot, not as the father of fractals, but as the instigator of a distinctive scientific posture. The guiding thread is methodological: accept the stubborn features of the data and use models as instruments for intuition rather than as axiomatic certificates of truth. In this perspective, scaling, intermittency and extremes are not peripheral imperfections around a well-behaved equilibrium; they are the very texture of economic and financial fluctuations. This naturally shifts attention from exogenous narratives to endogenous dynamics: interactions, feedback loops, and collective amplification mechanisms that can make systems intrinsically {\it fragile}. We argue that the importation of concepts from statistical physics -- criticality, disorder, emergence, multiplicative cascades -- should be read not as an artificial transposition but as a candid attempt to look for generic mechanisms compatible with empirical regularities observed across scales, from markets to macroeconomic aggregates. 
\end{abstract}
\onehalfspacing

\emph{Les paradoxes d’aujourd’hui
sont les préjugés de demain}

Marcel Proust, Les Plaisirs et les Jours (1896)  

\section*{Prologue: staring at data until it stares back}

Econophysics is sometimes caricatured as the naive transfer of physics tools into economics and finance. The caricature is convenient for both sides: economists can dismiss it as methodologically suspect, while physicists can pretend that their importation of concepts is automatically legitimate. The more faithful description is much simpler and, in fact, much more Mandelbrotian: \emph{econophysics is first and foremost a style of inquiry}. It privileges empirical regularities, insists on visual and qualitative confrontation with data, and is willing to use ``toy'' models---metaphors, analogies, minimal mechanisms---to organize robust facts before attempting any axiomatic closure.

This style was actually pioneered by Beno\^{i}t Mandelbrot. He did not merely propose a model for speculative prices; he opened a breach. The breach is epistemological: the claim that market data are \emph{not} well described by Gaussian fluctuations around an equilibrium; that large events are not peripheral anomalies but a central feature; and that scale invariance (or something close to it) is not an exotic curiosity but a plausible organizing principle.

Econophysics, in that sense, is the continuation of Mandelbrot's heresy by a statistical physics community that was itself going through a conceptual revolution. We learned, from critical phenomena, turbulence, spin glasses, chaos, and self-organized criticality, that complex collective behavior often cannot be inferred from individual ingredients. ``More is different'' is not a slogan; it is a methodological warning. Markets and economies---collections of interacting heterogeneous agents, feeding back on one another through prices, expectations, constraints---are precisely the kind of systems for which this warning is relevant.

The talk summarized here sketches a genealogical thread: from Mandelbrot's early work on speculative prices and fractional Brownian motion, to the data-driven discovery (or rediscovery) of fat tails and volatility clustering, to multifractality and multiplicative chaos, to practical implementations in option pricing and risk management, and finally to broader economic regularities such as Pareto and Zipf laws and the ``granularity'' hypothesis for macroeconomic volatility. The thread is not purely historical. It points to a persistent intellectual tension that has become clearer over the years: \emph{efficient} systems may also be \emph{fragile}, and endogenous dynamics can generate ``Black Swans'' even when exogenous news is scarce.

\section{What is econophysics? A methodological definition}

The term ``econophysics'' is not particularly important; what matters is the shift in attitude it signals (Bouchaud 2019). This section proposes a compact definition in four ingredients.

\begin{itemize}
    \item {(i) Priority to data; qualitative statistics; stylized facts}.
    
Econophysics begins with the empirical record. Not only with the mean and the variance, but with the full distribution, the temporal dependence structure, and the multi-scale organization of fluctuations. It tolerates, even encourages, ``eye-balling statistics'': stare at graphs and charts, compare scales, plot distributions on log-log axes, look for collapses, check robustness across assets, epochs, and sampling frequencies. In physics this is routine; in economics it has sometimes been regarded as indecorous, because it does not immediately speak the language of structural parameters.

Stylized facts are not a substitute for theory, but they are its constraints. They are also a defense against the tendency to overfit narratives to events. 

\item {(ii) Scale invariance, fat tails, concentration}

At the heart of the Mandelbrot program lies scale invariance and its consequences: power laws, heavy tails, and intermittency. ``Concentration'' is Mandelbrot's way of emphasizing that a small fraction of events can account for a large fraction of aggregate variation. In financial returns, a few extreme days can dominate long-run performance; in wealth, a small fraction of individuals hold a large fraction of total assets; in firm sizes, a small number of firms account for a large fraction of output.

Scale invariance is not an aesthetic preference. It is a statement about the absence of a characteristic scale, which typically indicates either (a) proximity to a critical point, (b) multiplicative growth and random proportional effects, or (c) aggregation of heterogeneous processes across scales. Each of these interpretations leads to different classes of models and different policy intuitions.

\item{(iii) Toy models, metaphors, intuitive mechanisms}

The econophysics style rejects the idea that a model must be derived from axioms of rationality to be legitimate. This is not necessarily an attack on rationality; it is rather an admission of epistemic modesty. We do not know what the correct axioms should be, and we know that many elegant axioms produce embarrassingly wrong predictions. Toy models are useful precisely because they clarify mechanisms and produce qualitative signatures that can be checked in data.

\item{(iv) A shared culture with statistical physics and complex systems}

The last ingredient is sociological and intellectual: econophysics is the application of the complex-systems mindset to economic data. Interactions matter, feedback loops matter, disorder matters, and collective phenomena can be qualitatively different from the behavior of the constituents. This culture was not invented in finance; it migrated there, with Mandelbrot as an early guide.
\end{itemize}

\section{A partial$^2$ timeline: Mandelbrot among the revolutions of statistical physics}

This section offers a ``partial$^{2}$'' timeline, which is an apt description: it is partial in selection and partial in the sense of being biased toward Mandelbrot and the physics community. But the point is not historiography; it is to show the remarkable synchronization between Mandelbrot's financial insights and the physics community's growing obsession with scaling, disorder, and emergence.

\subsection{1960s: speculative prices, Brownian heresy, and the fractal program}
Mandelbrot's 1963 paper on speculative prices is often presented as the birth certificate of the field. It attacked the Gaussian dogma by exhibiting heavy tails and proposing stable laws as a first approximation. Even when the pure L\'evy-stable hypothesis fails in detail (and it does), the core message survives: returns have far more probability mass in their tails than a Gaussian world allows.

By 1967, the conceptual apparatus broadens to include fractional Brownian motion (fBM): a Gaussian process with long-range dependence. One must be careful here. Financial returns themselves in liquid markets are close to uncorrelated (otherwise obvious arbitrages would exist), but \emph{volatility} is not. The long memory of amplitudes is the real empirical fact, and fBM is a natural object for thinking about scaling and memory.

\subsection{Late 1960s--1970s: critical phenomena, chaos, and early econophysics}
Wilson--Kadanoff scaling theory (1966/1970) is the canonical physics framework in which scale invariance becomes a predictive tool. The renormalization group teaches us that macroscopic laws can be insensitive to microscopic details, and that power laws can be universal. This intellectual background is essential: it makes it natural to look for universality in market fluctuations.

Around 1971, the Ruelle--Takens picture and the rise of chaos theory (and later Doyne Farmer's influence) emphasize that complex dynamics can arise endogenously in deterministic systems. The same year sees early econophysics-like papers (Weidlich, de Gennes), in parallel with Schelling's micromotives-macrobehavior. Again the same theme: aggregate patterns emerge from interactions, not from representative agents.

\subsection{1970s--1990s: turbulence, spin glasses, complex systems, SOC, Santa Fe, and the explosion}
In 1974 Mandelbrot returns to turbulence and multifractals, later formalized in the Frisch--Parisi multifractal framework (1980). Turbulence is the archetype of intermittent, multi-scale fluctuations; finance will later look like turbulence with a suit and tie.

In 1979 Parisi solves the ``spin glass'' problem (a milestone recognized by the 2021 Nobel Prize). The spin-glass paradigm---rugged landscapes, hierarchical organization, non-ergodicity, and the centrality of disorder---becomes part of the complex-systems toolbox, with a renewed interest (possibly defining a central paradigm) in the context of Machine Learning and AI. 

In 1987 Bak proposes the concept of self-organized criticality (SOC), popularizing the idea that power laws can arise without fine tuning. The same year, the first Santa Fe conference on the economy as a complex evolving system signals that complexity may be crucial to understand economic systems. By 1991, econophysics explodes both in academia and in the finance industry: Mantegna, Stanley, Farmer, Sornette, Galam, Zhang, and others, with parallel developments in quantitative firms (Prediction Company, CFM, etc.). In 1997, multifractal modeling of FX rates (Bacry--Muzy and collaborators) operationalizes the turbulence analogy proposed by Mandelbrot into concrete stochastic processes.

\section{Statistical ``phynance'': the initial ambition}

The first wave of econophysics in finance was not macroeconomics, not equilibrium theory, and certainly not welfare theorems. It was the humble  activity of \emph{measuring} before \emph{modelling}. I had proposed calling the field ``statistical phynance'' (after Alfred Jarry), but it did not stick.

\subsection{Fat tails: universal, but not L\'evy-stable}
A robust empirical fact is the presence of fat tails in return distributions. The word ``universal'' is tempting: across asset classes, one often finds tail exponents in a similar range, and one finds strong deviations from Gaussianity at short and medium horizons. But the data emphasizes an important caveat: empirical tails are fat, \emph{but not L\'evy-stable}. In practice, very large returns are less frequent than a pure stable law would predict; moments exist (at least up to some order); and there are cutoffs, regime shifts, and volatility states.

This nuance matters. A pure Lévy stable model is elegant but too rigid. The right attitude is to preserve the central insight---\emph{the market produces extremes endogenously and far more often than Gaussian theory predicts}---while allowing for more realistic mechanisms, such as stochastic volatility, multiplicative cascades, or mixtures of regimes.

\subsection{Volatility clustering and long memory: Mandelbrot's early insight}
Mandelbrot's famous sentence captures the central temporal regularity:
\begin{quote}
\emph{Large changes tend to be followed by large changes, of either sign, and small changes tend to be followed by small changes.}
\end{quote}
Returns have weak autocorrelation, but volatility does not. The autocorrelation of $|r_t|$ or $r_t^2$ decays slowly, often compatible with a power law. This is the financial analog of intermittency in turbulence: bursts of activity separated by calm periods, with structure across many time scales.

\subsection{Most ``jumps'' come from nowhere: excess volatility and endogenous dynamics}
Another striking empirical observation is that a large fraction of big price moves cannot be traced to identifiable public news. This is not a claim that news is irrelevant, but that the mapping between news and prices is weak, nonlinear, and mediated by market microstructure and crowd dynamics.

The classical economist may respond: ``news is often private or hard to measure.'' True, but the econophysicist retorts: then the relevant object is not news but the system's \emph{internal state}---liquidity, leverage, positioning, risk constraints, herding---because these govern the amplification of small perturbations into large moves. (And, by the way, the last thing an informed trader should do is to make prices jump!)

This connects to the theme of ``excess volatility'': price variability seems too large relative to what can be justified by changes in fundamentals. Whether one embraces that argument in its strongest form or not, the practical implication is the same: a large component of volatility is generated by the market itself.

\section{From turbulence to finance: multifractality and the Multifractal Random Walk}

The turbulence analogy is not merely rhetorical. It suggests a class of models in which variability is organized across scales through multiplicative mechanisms.

\subsection{Multifractality: many exponents, many scales}

A monofractal process is characterized by a single Hurst exponent; multifractality is characterized by a spectrum of scaling exponents. In finance, empirical structure functions often display anomalous scaling, suggesting intermittent cascades rather than simple Brownian scaling. Again, one must separate signal from noise: microstructure effects, finite samples, and non-stationarity can mimic multifractality. But the persistence of multi-scale volatility structure motivates explicit multifractal models.

\subsection{The Multifractal Random Walk: long memory in amplitudes}
The ``Multifractal Random Walk'' (MRW) is a famous scale-invariant time series relevant for turbulence and finance. The key idea is to model returns as
\[
r_t = \sigma_t \epsilon_t,
\]
with $\epsilon_t$ close to i.i.d.\ Gaussian, but with a stochastic volatility $\sigma_t$ whose logarithm has long-range correlations. This produces:
\begin{itemize}
  \item Heavy tails (mixture of Gaussians with random variance).
  \item Volatility clustering and long memory (through the correlation structure of $\log \sigma_t$).
  \item Approximate scaling and multifractal-like behavior.
\end{itemize}

Conceptually, MRW and related models shift the explanatory burden from returns to volatility. Prices may be close to martingales, but the \emph{activity level} of the market is not. It has its own dynamics, reminiscent of energy dissipation in turbulence.

\subsection{From physics to finance and back: cracks, landscapes, multiplicative chaos}
An important intellectual message is the two-way street: ideas migrate from physics to finance, but also from finance back to physics. Multifractal cracks (Vernede et al.) and hierarchical landscapes are examples where similar mathematics appears in different guises. Multiplicative chaos, for instance, is not a financial invention; it is a deep probabilistic construction (Kahane) that has become a central tool in modern probability and field theory (see Vargas \& Rhodes). Finance provides an empirical playground where log-correlated volatility fields and multiplicative cascades acquire a concrete operational meaning.

\section{From market data to applications: pricing, hedging, and synthetic series}

One should not judge econophysics only by its metaphors; one should also judge it by its usefulness for applications in option pricing, hedging, and portfolio construction when fat tails and volatility clustering are present.

\subsection{Fat tails and clustered volatility: the end of the Black--Scholes comfort zone}
In a Gaussian i.i.d.\ world, delta hedging errors are small and well behaved; risk can be summarized by variance; and option smiles should not exist. In real markets:
\begin{itemize}
  \item Volatility clustering makes risk state-dependent and persistent.
  \item Tail risk makes large hedging errors inevitable, leading to a demise of ``perfect replication'' and unambiguous option prices. 
  \item Implied volatility smiles are ubiquitous (and, in fact, already visible in Bachelier's empirical data in 1900!).
\end{itemize}
The last remark is not merely historical trivia. It reminds us that the empirical deviations from Gaussianity are old; what is new is the awakening to the true risk they entail and their industrial-scale quantification (Walter).

\subsection{Rough volatility: an fBM extension of MRW}
``Rough volatility'' can be seen as a modern synthesis of Mandelbrot's fBM intuition with the stochastic-volatility viewpoint. Empirically, volatility appears \emph{rougher} than Brownian motion, with an effective Hurst exponent $H<1/2$ at short time scales. (The Multifractal Random Walk corresponds formally to the limit $H \to 0$). This roughness is not a small detail; it changes the term structure of implied volatility skews, the behavior of short-dated options, and the statistics of realized variance.

In the Mandelbrot genealogy, rough volatility is a return of fBM through the back door: not for returns themselves (which remain close to uncorrelated), but for volatility as a rough, persistent field. This is perhaps one of the clearest examples where an idea that looked like an academic curiosity becomes a practical modeling tool.

\subsection{Wavelets and multi-scaling: generating high-quality synthetic time series}
Recent work using wavelets (with R. Morel \& S. Mallat) as a method to capture multi-scaling and generate synthetic financial time series. Wavelets are naturally adapted to multi-resolution structure; they provide localized scale-by-scale representations. For finance, the goal of high-quality synthetic series is not entertainment: it is stress testing, benchmarking of strategies, and risk management under realistic temporal dependence structures.

\section{From financial markets to economics: power laws everywhere}

 The striking observation is the pervasiveness of power laws in socio-economic data, from finance to economics at large, many known to Mandelbrot.

\subsection{Pareto wealth tails and toy models}
The Pareto distribution of wealth is a canonical example of fat tails: a small fraction holds a disproportionate share. This fact has spawned a host of toy models, many of them ``econophysics'' flavored, often based on random exchange, multiplicative growth, saving propensities, or preferential attachment. The key is not to claim that any one toy model is ``the'' economy, but to identify mechanisms that robustly generate Pareto tails under broad conditions (Bouchaud, Mézard).

\subsection{Zipf laws for firms, cities, sectors}
Zipf's law---roughly, rank proportional to size$^{-1}$---appears in firm sizes, city sizes, and sector sizes. Such regularities suggest underlying proportional growth mechanisms or equilibrium between growth and fragmentation. Again, one must avoid overinterpretation: institutions and policies must matter. But the persistence of Zipf-like laws across countries indicates that strong statistical regularities can coexist with institutional diversity. Again, random multiplicative growth models are useful to think about such regularities (Gabaix 1999).

\subsection{Granularity and the failure of the Central Limit Theorem at the macro level}
Perhaps the most policy-relevant part of this zoom-out is the ``granularity hypothesis'' (Gabaix): if the distribution of firm sizes is fat-tailed, then the idiosyncratic shocks of the largest firms do not average out. The Central Limit Theorem fails in practice because the effective number of independent contributors is small: the top firms dominate.

The crucial empirical finding is that growth rate fluctuations decay \emph{much slower} than $S^{-1/2}$ with firm size $S$. This slow decay is directly connected to the ``small shocks / large business cycle puzzle'': macroeconomic volatility appears too large to be explained by the aggregation of many independent micro shocks in a world where everything averages out.

The granularity view is a clean, falsifiable mechanism: measure the firm size distribution, quantify the contribution of the top firms, and assess whether macro fluctuations can indeed be traced to micro-level granularity. It does not claim that this is the whole story. Financial amplification, network effects, leverage cycles, and policy feedbacks also matter. But it provides an explicit route by which ``micro'' noise can remain ``macro''.

\section{Looking for mechanisms: why do small causes lead to large effects?}

At this point, the intellectual program anticipated by Mandelbrot becomes explicit: we observe fat tails, clustering, intermittency, large crises with weak news triggers, power laws in wealth and firm sizes, and anomalously large macro volatility. What mechanisms can plausibly generate these patterns?

\subsection{Fluctuations + interactions $\Rightarrow$ emergence and endogenous dynamics}
Statistical physics offers a template: fluctuations are amplified by interactions. When local variables are coupled, the system can develop collective modes, long correlations, and nontrivial responses. The list of examples one can think of---earthquakes, avalanches, epileptic seizures, financial and economic crises, fads and fashions---is intentionally heterogeneous. The point is that large events can be \emph{endogenous} outcomes of systems with threshold dynamics, feedback, and slow driving.

In finance, interaction takes many forms: order flow impacts prices; price moves affect risk constraints; risk constraints feed back on trading; expectations are formed from past prices; and liquidity itself is state-dependent. Even if each agent is ``rational'' in a narrow sense, the aggregate can be unstable.

\subsection{Self-organized criticality and the efficiency--fragility tradeoff}
SOC (Bak) provides an overarching metaphor: systems driven slowly with local relaxation can self-tune near criticality, producing scale-free avalanches. Whether financial markets are literally SOC systems is debatable, but the metaphor captures an uncomfortable possibility: markets may evolve toward states that are locally efficient---liquidity provision, tight spreads, high leverage, intense competition---yet globally fragile. Efficiency at the micro level may create critical sensitivity at the macro level (Bouchaud 2024 and refs. therein).

This perspective complements (and challenges) the standard efficient market hypothesis. A market can be informationally efficient in some sense, while being dynamically unstable because of feedback and endogenous risk-taking.

\subsection{Multiplicative growth and fat tails}
As already mentioned, multiplicative processes---random proportional growth---are another robust route to fat tails. They appear in wealth accumulation, firm growth, and cascade models of volatility. In finance, multiplicative volatility cascades naturally generate heavy tails without requiring explicit ``news jumps.'' In economics, multiplicative growth under constraints and entry/exit can generate Pareto and Zipf laws.

It is tempting to suggest that SOC and multiplicative growth may be overarching mechanisms for Black Swans. One should not seek a single universal explanation, but these two families of mechanisms have the virtue of being both mathematically generative of power laws and conceptually aligned with the observed endogeneity of large events.

\section{Mandelbrot's intellectual legacy: ``Les paradoxes d'aujourd'hui sont les pr\'ejug\'es de demain''}

When thinking about Mandelbrot's legacy, it is apt to quote Proust: ``Today's paradoxes are tomorrow's prejudices.'' His claims were paradoxical in a Gaussian, equilibrium-based intellectual climate: heavy tails, scale invariance, discontinuous-looking moves, long memory in volatility. Many of these claims are now banal. Option traders live with smiles; risk managers live with fat tails; regulators live with systemic crises.

But the Proust quote is also a warning. Once a paradox becomes a prejudice, it risks becoming another dogma. The current fashion for power laws, multifractals, or rough volatility can become superficial if it is not disciplined by empirical rigor and mechanistic clarity. The Mandelbrot legacy should not be the worship of fractals but rather the insistence that we must build models that respect the data, that we should not be embarrassed by complexity, and that endogenous crises are not an afterthought but a central object of study.

The econophysics program remains enticing because the most worrisome phenomena---systemic crises in finance and macroeconomics---still lack a fully satisfactory theory. And perhaps they will always resist complete domestication, for the same reason fluid turbulence resists it: multi-scale, intermittent, feedback-driven dynamics can be characterized statistically, modeled approximately, and managed pragmatically, but not reduced to a single closed-form truth.

\newpage
\section*{References (selected)}

\end{document}